\documentclass[useAMS,usenatbib]{mn2e}
 \usepackage{times,graphicx,aas_macros}

\begin{document}

\title[CO\ {\it J}(1$\rightarrow$0)\ observations of cluster
starbursts]{CO interferometry of gas-rich spiral galaxies in the
  outskirts of an intermediate redshift cluster} \author[J. E. Geach
et al.]  {\parbox[h]{\textwidth}{ James\ E.\ Geach,$^1$\thanks{E-mail:
      j.e.geach@durham.ac.uk} Ian\ Smail,$^1$
    Kristen\ Coppin,$^1$  Sean\ M.\ Moran,$^2$ Alastair\ C.\ Edge,$^1$\\
    and Richard~S.~Ellis$^{3,4}$ }
  \vspace*{6pt}\\
\noindent $^1$Institute for Computational Cosmology, Durham University, South Road, 
Durham. DH1 3LE. UK.\\
\noindent $^2$Department of Physics and Astronomy, Johns Hopkins University, 3400 N Charles St, Baltimore, MD 21218 \\
\noindent $^3$Department of
Astrophysics, University of Oxford, Keble Road, Oxford, OX1 3RH, UK.\\
\noindent $^4$California Institute of Technology, 1200 East
California Boulevard, Pasadena, CA 91125 }

\date{}

\pagerange{\pageref{firstpage}--\pageref{lastpage}} \pubyear{2009}

\maketitle

\label{firstpage}

\begin{abstract}We present IRAM Plateau de Bure Interferometer 3\,mm
  observations of CO\ {\it J}(1$\rightarrow$0) emission in two
  24$\mu$m-selected starburst galaxies in the outskirts
  ($\sim$2--3$R_{\rm virial}$) of the rich cluster Cl\,0024+16
  ($z=0.395$). The galaxies' inferred far-infrared luminosities place
  them in the luminous infrared galaxy class (LIRGs, $L_{\rm
    FIR}>10^{11}L_\odot$), with star formation rates of
  $\sim$60\,$M_\odot$\,yr$^{-1}$. Strong CO\ {\it J}(1$\rightarrow$0)
  emission is detected in both galaxies, and we use the CO line
  luminosity to estimate the mass of cold molecular gas, $M({\rm
    H_2})$. Assuming $M({\rm H_2})/L'_{\rm CO} = 0.8\,M_\odot({\rm
    K\,km^{-1}\,pc^2})^{-1}$, we estimate $M({\rm H_2}) =
  (5.4$--$9.1)\times10^9M_\odot$ for the two galaxies. We estimate the
  galaxies' dynamical masses from their CO line-widths, $M_{\rm
    dyn}\sim1$--$3\times10^{10}M_\odot$, implying large cold gas
  fractions in the galaxies' central regions.  At their current rates
  they will complete the assembly of $M_\star\sim10^{10}M_\odot$ and
  double their stellar mass within as little as $\sim$150\,Myr. If
  these galaxies are destined to evolve into S0s, then the short
  time-scale for stellar mass assembly implies that their major
  episode of bulge growth occurs while they are still in the cluster
  outskirts, long before they reach the core regions. Subsequent
  fading of the disc component relative to the stellar bulge after the
  gas reservoirs have been exhausted could complete the transformation
  of spiral-to-S0.
\end{abstract}
\begin{keywords}
  clusters: galaxies, clusters: individual: Cl\,0024+16, galaxies:
  starburst, evolution
\end{keywords}

\begin{table*}
  \caption{Details of the two galaxies in our study:
    MIPS\,J002621.7+171925.7 and  MIPS\,J002721.0+165947.3 (Geach et
    al.\ 2009). We present the results of
    the CO observations: CO(1--0) line widths, luminosities and
    corresponding H$_2$ gas mass. Mid-infrared
    observations are from Geach et al.\ (2006 \& 2009), and the
    spectroscopic redshifts are from Czoske
    et al.\ (2001) and derived from the [O~{\sc ii}] emission.  Note the double Gaussian
    profile of MIPS\,J002721.0 was a four parameter fit, with a single
    FWHM and amplitude and two velocity offsets.}
\begin{tabular}{@{\extracolsep{\fill}}lcccccccccc}
  \hline
  Target & $\alpha_{\rm J2000}$ & $\delta_{\rm J2000}$ & $z$  &
  $L_{\rm FIR}$ &  SFR & $V_{\rm FWHM}$& $\Delta v^\dagger$ &  $L'_{\rm CO}$ & $M({\rm H_2})^{\dagger\dagger}$~~\cr 
  & (h\ m\ s) & ($\circ$\ $'$\ $''$) &  &
  ($10^{11}L_\odot$)  & ($M_\odot$\,yr$^{-1}$) & (km\,s$^{-1}$) &
  (km\,s$^{-1}$) & (10$^{10}$\ K\ km\ s$^{-1}$\ pc$^2$)  & ($10^{9}M_\odot$) \cr
  \hline
  MIPS\,J002621.7 & 00\ 26\ 21.7 & +17\ 19\ 26.4 & 0.3803  & $3.1\pm0.2$ &
  $56\pm16$ & 144$\pm$13 & $5$ & $0.68\pm0.06$ & $5.4\pm0.5$ \cr
  MIPS\,J002721.0 &  00\ 27\ 21.1 &  +16\ 59\ 49.9 & 0.3964  & $3.2\pm0.2$ &
  $59\pm16$  & 158$\pm$34 & $-150$, $30$   & $1.14\pm0.11$ &
  $9.1\pm0.9$\cr
  \hline
  \multicolumn{10}{l}{$\dagger$\,Offset from [O~{\sc ii}]
    line. MIPS\,J002721.0 is best fit with a double Gaussian, and we
    quote the offsets of both peaks. The uncertainty is $\sim$30\,km\,s$^{-1}$.}\cr
  \multicolumn{10}{l}{$\dagger\dagger$\,Assuming a CO luminosity to
    total gas mass
    conversion factor of $\alpha = 0.8\,M_\odot({\rm
      K\,km^{-1}\,pc^2})^{-1}$}
\end{tabular}
\end{table*}

\section{Introduction}

One key requirement for a starburst is the presence of a reservoir of
dense, cold gas that can be efficiently converted to stars. For
galaxies entering rich clusters this is especially important, because
they are expected to be affected by mechanisms that can remove cold
gas from the haloes and discs of infalling galaxies (e.g.\
ram-pressure stripping, Gunn \& Gott\ 1972) or prevent further cooling
of gas within galaxies' dark matter halos (starvation or
strangulation, e.g.\ Larson, Tinsley \& Caldwell\ 1980; Bekki, Couch
\& Shioya\ 2002). This environmental dependence has a profound
influence on cluster galaxies' evolutionary histories, the net effect
of which is the eventual termination of star formation (there is
virtually no residual star formation in the cores of local
clusters). The observational evidence for the gradual truncation of
star formation in clusters is the conspicuous disappearance of
star-forming disc galaxies in the cores of rich clusters since
$z\sim1$. It has been proposed that the spiral galaxies in distant
clusters must be transforming into passive lenticular (S0) galaxies,
since the fraction of S0s in rich clusters is believed to increase at
the same time that the spirals begin to vanish (Dressler et al.\
1997). However, the detailed nature of the physics controlling this
evolution is still poorly understood. For example, what process is
responsible for the transformation of the bulge-to-disc ratio of the
spiral population? (Kodama \& Smail 2001).

A growing body of observational evidence suggests that distant
clusters contain significant populations of dust obscured starburst
galaxies in their peripheral regions (e.g.\ Duc et al.\ 2000; Biviano
et al.\ 2004; Coia et al.\ 2004; Geach et al.\ 2006; Marcillac et al.\
2007; Bai et al.\ 2007; Dressler et al.\ 2008; Koyama et al.\ 2008;
Fadda et al.\ 2008). Geach et al.\ (2009) proposed that the population
of luminous infrared galaxies (LIRGs) residing in the `infall'
population of $z\sim0.5$ clusters could be examples of spiral galaxies
undergoing an episode of bulge growth via circumnuclear starburst, and
therefore excellent candidates for the progenitors of local massive
S0s.  Geach et al.\ (2009) estimate that, given the cluster LIRGs'
current SFRs, 10$^{10}M_\odot$ of stars could be built up within a few
100\,Myr. Although this simple model suggests that these galaxies
could evolve onto the bright end of the local cluster S0 luminosity
function by $z=0$, we still lack crucial observational constraints,
such as the likely duration of the starburst, where it occurs in the
cluster, and limits on the final stellar mass of the bulge.

In this Letter we refine our model by asking a simple question: what
is the mass of the cold molecular gas reservoir in these starbursts?
Using the IRAM Plateau de Bure Interferometer we have conducted a
pilot study of the cold gas properties of two dusty starburst galaxies
in the distant cluster Cl\,0024+16. The initial results are presented
here. Throughout we assume $h=0.7$ in units of
100\,km\,s$^{-1}$\,Mpc$^{-1}$, $\Omega_\Lambda = 0.7$ and $\Omega_{\rm
  m} = 0.3$. The projected scale at $z=0.395$ in this model is
5.3\,kpc/$''$, and the luminosity distance is $D_L\sim2140$\,Mpc.

\section{Observations}

The two galaxies chosen for this pilot study were selected as two of
the brightest in our {\it Spitzer Space Telescope} InfraRed
Spectrograph (IRS) survey of 24$\mu$m-selected members of Cl\,0024+16
($z=0.395$). Full details of the mid-infrared observations can be
found in Geach et al.\ (2006 \& 2009). Both galaxies have
$S_{24}>0.9$\,mJy, and implied far-infrared luminosities $L_{\rm FIR}
> 10^{11}L_\odot$; they are in the LIRG class. The far-infrared
luminosity is estimated from the luminosity of the 7.7$\mu$m
polycyclic aromatic hydrocarbon (PAH) emission line, extrapolating
from the tight correlation between PAH strength and total infrared
luminosity observed in local star forming galaxies (Smith et al.\
2007; Geach et al.\ 2009).  Both galaxies are in the outskirts, or
infall region, of the cluster: they large clustocentric radius,
$1.9R_V$ and $2.8R_V$, where $R_V$ is the virial radius,
$\sim$$1.7$\,Mpc (Treu et al.\ 2003). Further details of the two
galaxies is provided in Table\ 1.

The observations were conducted using the IRAM Plateau de Bure
Interferometer on 31 July 2008 (MIPS\,J002621.7) and 25 August 2008
(MIPS\,J002721.0) as part of program S035. The exposure times were
9.5\,hr and 8.2\,hr for each source respectively using 5 antennae. The
observing conditions were excellent in terms of atmospheric phase
stability, however any anomalous and high phase-noise visibilities
were flagged. Data were calibrated, mapped and analysed using the IRAM
{\sc gildas} software (Guilloteau \& Lucas\ 2000).  Phase and flux
calibration was performed using 3C454.3 and 0007+106. Secondary flux
calibrators also included the sources MWC349, 3C454.3, 0119+115, 3C84,
2145+067.  We targeted the CO(1--0) 115.27\,GHz rotational transition,
which at $z=0.4$ is redshifted into the 3\,mm band, with $\nu_{\rm
  obs} = 82.63$\,GHz. The central frequency of the 3\,mm receiver was
set to the frequency of the redshifted CO(1--0) line at the systemic
redshift of each galaxy derived from optical spectroscopy (Table\
1). The correlator was set-up with 2.5\,MHz spacing (2$\times$64
channels, 320\,MHz bandwidth) to comfortably cover the CO line should
it be offset from the systemic redshift, and to take into account a
broad velocity profile. The noise per 18\,km\,s$^{-1}$ channel for the
two observations is 1.4\,mJy\ beam$^{-1}$ and 1.3\,mJy\ beam$^{-1}$.

\begin{figure*}
\includegraphics[width=0.49\textwidth]{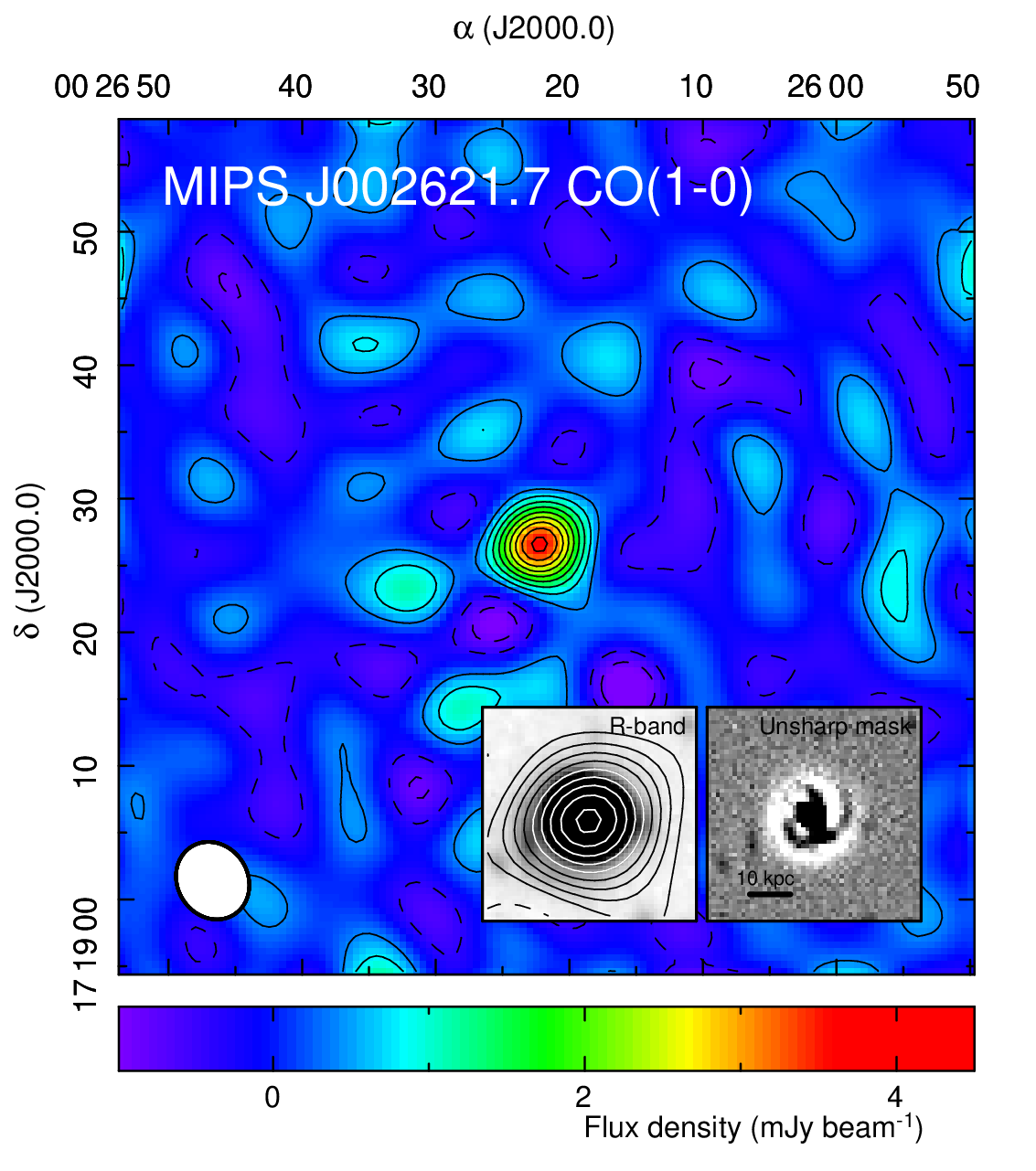}\includegraphics[width=0.49\textwidth]{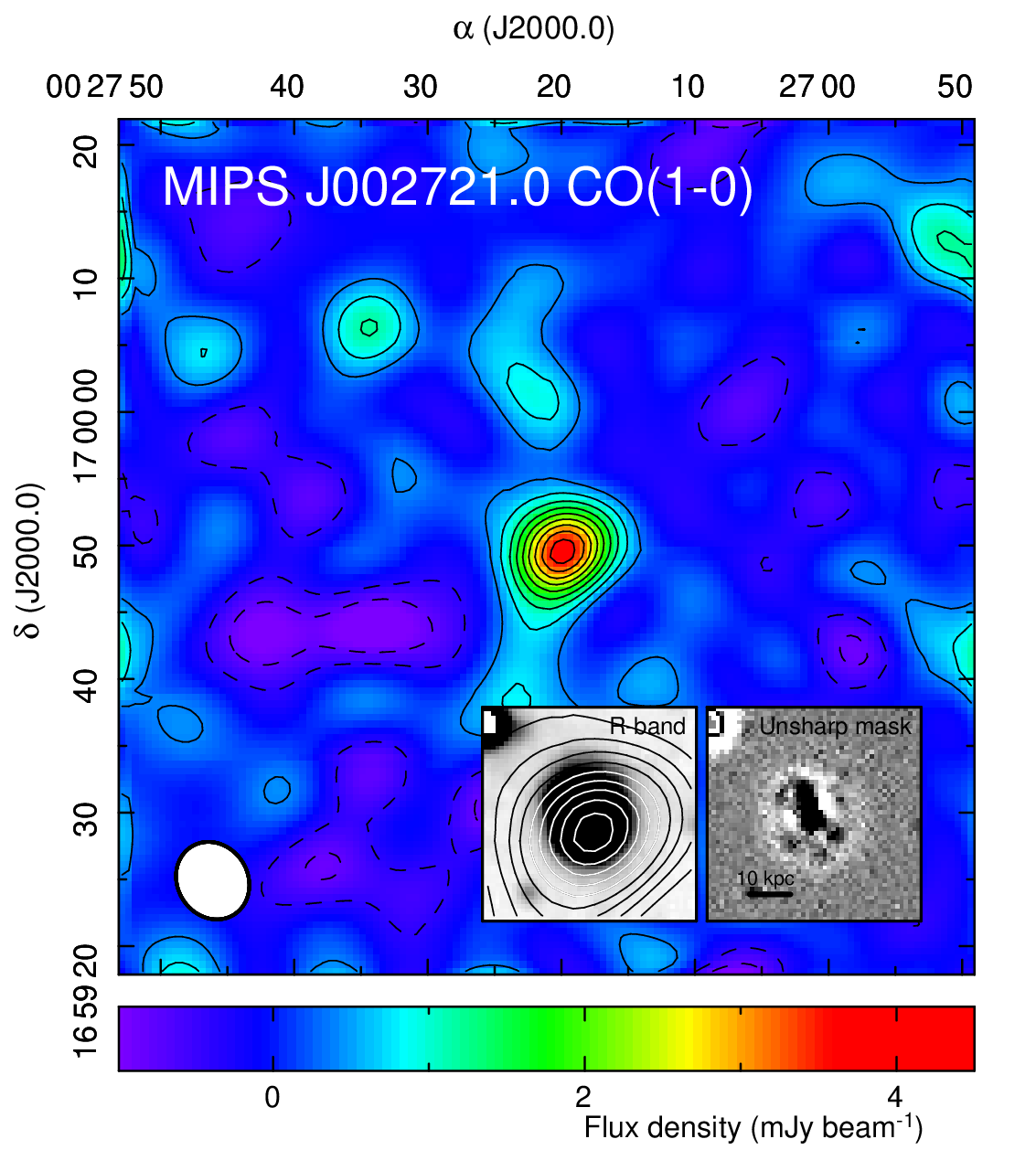}
\caption{Velocity integrated, clean CO(1--0) emission maps of the two
  galaxies in this study; each map is integrated over
  210\,km\,s$^{-1}$. The size and shape of the beam is indicated by
  the white ellipse, and all images are orientated North top, East
  left. Contours are shown at integer multiples of the typical
  r.m.s. noise, $\sim$0.4\,mJy, in each map (negative contours down to
  $-$2$\sigma$ are identified with dashed lines). The inset images are
  $10\times10$\,arcsec cut-outs of a Subaru {\it R}-band image (Kodama
  et al.\ 2006) around the position of the galaxy. We present two
  versions of the thumbnail: the standard grey-scale luminosity
  (overlaid with CO(1--0) contours at the same levels as the main
  map), and an un-sharp masked version. The un-sharp masked image is
  made by subtracting a Gaussian kernel smoothed version of the {\it
    R}-band image (kernel FWHM corresponds to 1\,kpc) from the
  un-processed version and reveals some of the small scale structure
  not visible in the original image. Note that MIPS\,J002621.7 appears
  to be a near face-on spiral and MIPS\,J002721.0 exhibits either an
  edge-on disc structure, or is made up from two components separated
  on scales of $\sim$10\,kpc.}
\end{figure*}

\section{Results \& Discussions}

\subsection{CO(1--0) line luminosity and H$_2$ gas mass}

We detect CO(1--0) emission in both galaxies with significances of
$9.3\sigma$ and $9.5\sigma$ respectively. Note that these observations
represent the highest redshift CO detection in un-lensed LIRGs yet
achieved (see Melchior \& Combes\ 2008). In Figure\,1 we present
velocity integrated emission maps, and the spectra extracted from the
peak pixel of these are shown in Figure\,2.  We estimate the line
luminosities by fitting the spectra with Gaussian profiles, although
in the case of MIPS\,J002721.0, the broad line is better fit with a
double Gaussian (there is a marginal improvement on the $\chi^2$
fit). We discuss the physical interpretation of the line profiles in
\S3.2.

We calculate $L'_{\rm CO}$ (in the usual units of
K\,km\,s$^{-1}$\,pc$^2$) from the integrated line emission according
to e.g. Solomon \& Vanden\ Bout\ (2005): $ L'_{\rm CO} =
{3.25\times10^7S_{\rm CO}\Delta v D_L^2}{\nu_{\rm obs}^{-2}
  (1+z)^{-3}}$. The line luminosities are given in Table\,1 with
1-$\sigma$ errors estimated by repeatedly re-evaluating the line fits
after adding noise to the spectra. The `noise' is randomly drawn from
a Gaussian distribution with a width equivalent to the variance in the
data sampled in the extreme wings ($|V_{\rm LSR}| >
500$\,km\,s$^{-1}$) of the line. We take the standard deviation in the
resulting $L'_{\rm CO}$ values from 1000 of these `bootstrap'
realisations as the 1$\sigma$ uncertainty in the measured value.

It is assumed that the CO luminosity linearly traces the molecular gas
mass: $M({\rm H_2}) = \alpha L'_{\rm CO}$ (e.g.\ Young \& Scoville
1991; Solomon \& Vanden Bout\ 2005) with the conversion factor
$\alpha$ typically taken as either $\alpha=4.6\,M_\odot({\rm
  K\,km^{-1}\,pc^2})^{-1}$ for `normal' spirals (the Galactic
conversion factor), or $\alpha=0.8\,M_\odot({\rm
  K\,km^{-1}\,pc^2})^{-1}$ for ultraluminous infrared galaxies
(ULIRGs, $L_{\rm FIR}>10^{12}L_\odot$).  The cluster starbursts are
not quite in the ULIRG class, but are at the extreme end of what would
be considered a `normal' IR-bright galaxy (see Gao \& Solomon\
2004). We therefore adopt the more conservative ULIRG conversion when
estimating the H$_2$ gas mass (Table\ 1), and we discuss the
implications of this in \S3.4. In this sense, the estimated H$_2$
masses could be considered lower limits.

\subsection{Morphology, geometry \& dynamics}

Deep, high-resolution (FWHM$\sim$0.8$''$) optical imaging of these
galaxies (Fig.~1, inset) allows us to constrain their
orientation\footnote{Unfortunately both galaxies are outside the
  sparse {\it HST} WFPC2 mosaic of this cluster (Treu et al.\ 2003)}.
MIPS\,J002621.7 appears to be nearly face-on to the line of sight,
with a bright core and distinctive arms. MIPS\,J002721.0 is slightly
more complicated; an un-sharp masked {\it R}-band image reveals a ring
of bright knots and a bright core or bar that appears to be slightly
extended to the north-east. It could also be orientated close to
face-on.

The CO(1--0) spectrum for MIPS\,J002721.0 exhibits a double peaked
line profile (Fig.~2), and we show in Figure~3 that the peak of the CO
emission traced by this velocity shear describes a locus that runs
$\sim$10\,kpc across the direction of the extended optical
emission. This could be interpreted as a rotating disc, with an
angular separation between the velocity peaks of $(1.8\pm0.8)''$ or
$(9.5\pm4.2)$\,kpc. Note that the peak of the CO emission at the
[O~{\sc ii}] redshift is centred on the bright central core.  An
alternative situation is that this galaxy is in the process of a
merger along the line-of-sight, and the CO(1--0) emission profile
originates from emission from two gas-rich clumps.

\begin{figure}
\centerline{\includegraphics[width=0.49\textwidth]{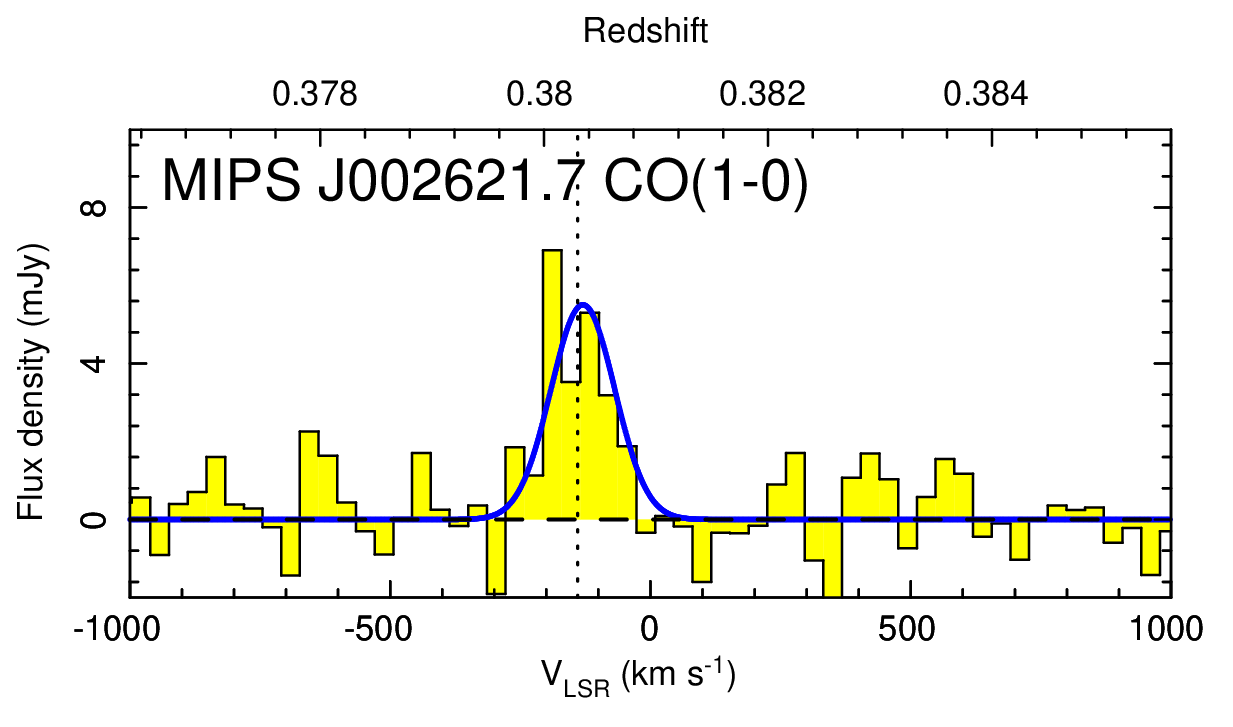}}
\centerline{\includegraphics[width=0.49\textwidth]{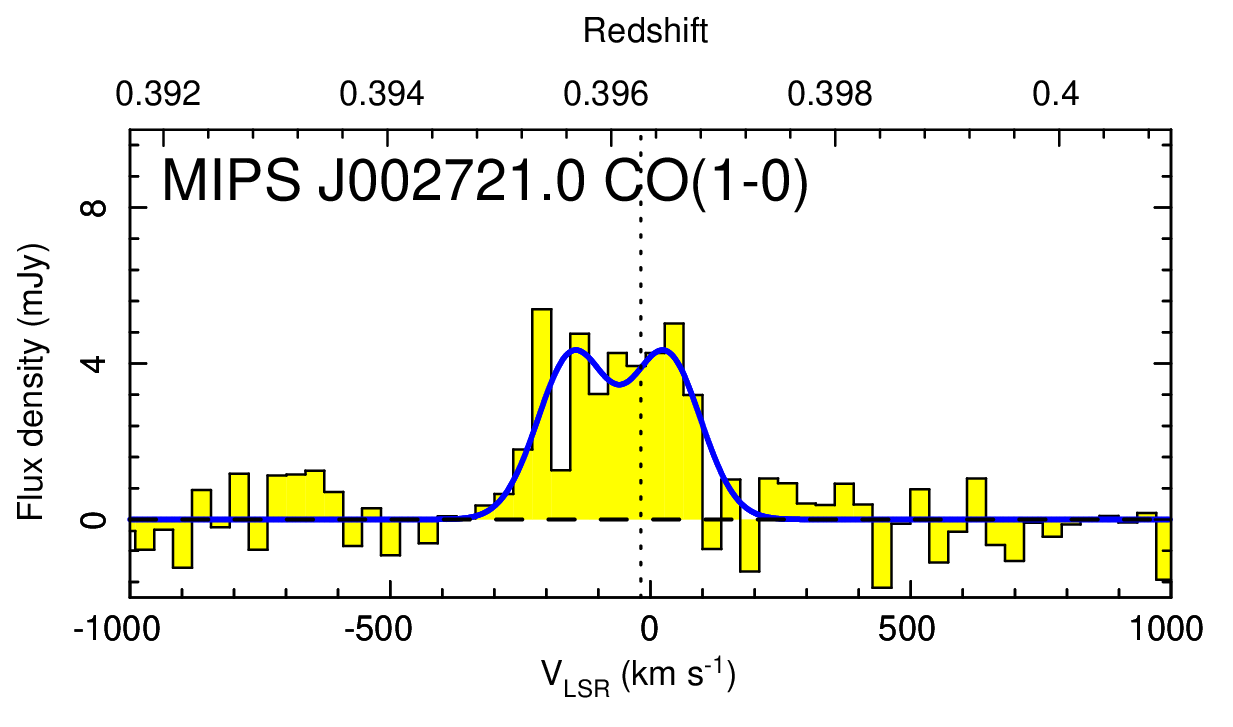}}
\caption{Millimeter spectra of the galaxies. The redshift of the
  target galaxy derived from their [O~{\sc ii}]$\lambda$3727 emission
  line is indicated by the vertical dashed lines, and expected to be
  accurate to $\sim$30\,km\,s$^{-1}$. The systemic CO redshifts agree
  with the optical redshifts (Czoske et al.\ 2001; Moran et al.\
  2006), although in the case of MIPS\,J002721.0 the broad
  blue-shifted profile is best fit by a double Gaussian with a peak
  separation of $(180\pm50)$\,km\,s$^{-1}$. This could be evidence of
  a rotating disc, or could also be interpreted as a merger. Note that
  redshifts have been corrected for heliocentric velocity (see
  Table~1).}
\end{figure}

Finally, we can use the CO line widths to estimate the galaxies'
dynamical masses. For a rotating disc of radius $R$, the dynamical
mass enclosed within $R$ is $M_{\rm dyn}\sin^2i=4\times10^4RV_{\rm
  FWHM}^2$ (Neri\ et al.\ 2003). One way of estimating the likely size
of the molecular disc is to force these galaxies to fall on the global
Schmidt-Kennicutt relation, connecting star formation surface density
to the gas surface density: $\Sigma_{\rm SF} \propto \Sigma_{\rm
  H_2}^{1.4}$ (Schmidt\ 1959; Kennicutt\ 1998). Assuming the gas
masses and SFRs given in Table\,1, this implies discs of radius
$\sim$0.4\,kpc and $\sim$1\,kpc for MIPS\,J002621.7 and
MIPS\,J002721.0 respectively. These values are not unreasonable, given
that observations of local LIRGs indicate large concentrations of gas
in their inner few kpc (Sanders \& Mirabel\ 1996). However, clearly
the latter value is at odds with the $\sim$10\,kpc separation of the
velocity peaks in MIPS\,J002721.0, and of course, these radii would
further increase if we had adopted a higher CO-to-H$_2$ mass
conversion. Assuming $i=10^\circ$ for MIPS\,J002621.7, we find $M_{\rm
  dyn}=(1.1\pm0.2)\times10^{10}M_\odot$. This ignores the systematic
uncertainty on the inclination angle, which would clearly affect these
mass estimates.  The orientation of MIPS\,J002721.0 is similarly
uncertain -- it could be a face-on merger, or an inclined
disc. Assuming it is a disc of radius 5\,kpc, with $i=30^\circ$
(appropriate for a random orientation), this yields $M_{\rm
  dyn}=(2.9\pm0.9)\times10^{10}M_\odot$. In a merger model this
estimate would increase by a factor $\sim$2 (Neri et al.\ 2003; Genzel
et al.\ 2003).

\subsection{Star formation efficiency \& timescales}

If we assume that the far-infrared luminosity is dominated by dusty
star-formation, then the star formation efficiency can be represented
by $L_{\rm FIR}/L'_{\rm CO}$. For these galaxies, we have $L_{\rm
  FIR}/L'_{\rm CO} = 46\pm2$ and $L_{\rm FIR}/L'_{\rm CO} =
28\pm2$. This is in the range observed in local ($z<0.1$) spirals,
which have $L_{\rm FIR}/L'_{\rm CO} <100$ (Solomon \& Vanden Bout\
2005). In contrast, much larger values of $L_{\rm FIR}/L'_{\rm CO} >
100$ are observed in ULIRGs and SMGs over a wide range of redshift
($0.1<z<3$) implying higher SFEs, perhaps driven by more violent
mergers and interactions (Downes, Solomon, \& Radford\ 1993; Tacconi
et al.\ 2008; Coppin et al.\ 2008). 

Recently, Daddi et al.\ (2008) have presented evidence for similar
`low-efficiency' star formation in two {\it BzK}-selected galaxies at
$z\sim1.5$. Those authors argue that while the most luminous galaxies
such as ULIRGs and SMGs are indeed special cases of intense,
high-efficiency star-formation, the {\it general} case could be a less
efficient (yet still vigorous) mode of star-formation fuelled by a
large reservoir of gas. Unfortunately a lack of CO observations of
LIRGs at intermediate redshift makes such comparisons challenging.
More extensive surveys in the ALMA era will no-doubt dramatically
improve our understanding of this issue.

Given our measured gas mass, we can estimate the gas depletion
timescale from the galaxies' current SFRs (Table\ 1). Assuming the
activity continues at the observed rate, and that all the gas is
converted to stars, the lifetime of the starbursts are $\sim$100\,Myr
and $\sim$160\,Myr for MIPS\,J002621.7 and MIPS\,J002721.0
respectively. Note that this is much shorter than the time it takes
for a galaxy on a radial orbit to traverse the distance from the
outskirts of the cluster to the core, which is of order 2--5\,Gyr for
Cl\,0024+16 (Treu\ et\ al.\ 2003). The implication of these short
time-scales is that self-exhaustion of the gas-reservoirs
dominates these galaxies' star formation histories. Moreover it
is only the environment within a few 100\,kpc of the cluster core that
can ram-pressure strip the disc gas (and thus truncate star
formation). The major episode of star-formation for these very active
galaxies will be long-since over by the time the galaxies reach the
core. Note that this may not be true of the `general' star forming
population, where lower SFRs mean that gas is not exhausted in such a
dramatic way, and allows star-formation in less active systems to
endure further into the cluster environment (e.g.\ Moran et al.\
2007).

\subsection{Ordinary, extraordinary or somewhere in-between?}

We have treated these galaxies as ULIRGs in terms of H$_2$ mass
estimation, but how would our conclusions change if we applied the
Galactic conversion of $\alpha=4.6$? The obvious implication is that
our derived gas masses would increase by a factor 6$\times$, and so
the galaxies' descendants could have stellar masses of order
$\sim$10$^{11}M_\odot$. If the majority of this additional mass is
built in the bulge region, then this further supports a model where
spiral galaxies in distant clusters can evolve into some of the most
massive S0 galaxies in the cores of local clusters. The longer
gas-depletion timescales ($\sim$500\,Myr and $\sim$900\,Myr) are still
shorter than the time it takes the galaxies to reach the virial radius
in Cl\,0024+16; our conclusion that the majority of the bulge growth
occurs in the outskirts of the cluster still holds.

These galaxies are clearly different in nature to the Milky Way --
they are forming stars at $\sim$15$\times$ the Galactic rate for
example. Although these galaxies are not as extreme as ULIRGs, it is
likely that their mode of star formation is more similar to ULIRGs
than the `quiescent' mode of star formation that occurs in the discs
of local spirals. It could be driven by a circumnuclear mode where
cold gas is being funnelled into the central regions for
example. Taking this into account, we assume that the ULIRG conversion
factor is most appropriate here.

\begin{figure}
\begin{center}\includegraphics[width=0.49\textwidth]{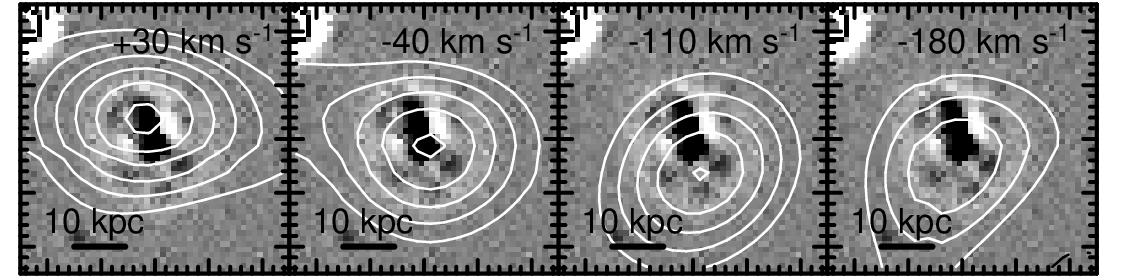}\end{center}
\caption{Un-sharp masked $10''\times10''$ {\it R}-band images of
  MIPS\,J002721.0. The main morphological details are a ring of knots
  and a central bright core which is extended to the north-east. The
  four images are overlaid with the CO(1--0) emission slices taken at
  70\,km\,s$^{-1}$ intervals. The change in peak position represents
  the velocity shear across the galaxy. The contours are taken from
  the cleaned map, and are at integer multiples of the noise,
  $\sim$0.65\,mJy\,beam$^{-1}$. The separation of the velocity peaks
  correspond to $(9.5\pm4.2)$\,kpc.}
\end{figure}

\subsection{The cluster environment and the starbursts' fate}

This pilot study is not extensive enough to make strong conclusions
about the potential role of the cluster environment on the gas
properties of these galaxies; a more complete survey is
needed. However we can make some comments about the possible
mechanisms that might be operating, the likely fate of these galaxies,
and the implications this has for models of galaxy evolution in rich
clusters.

At their large clustocentric radii, these galaxies are not affected by
ram-pressure stripping (although weak pressure of the ICM on the discs
of the galaxies could promote star formation in the disc; see Bekki \&
Couch\ 2003 and the discussion in Geach\ et\ al.\ 2009). Loosely bound
gas in the halos of the galaxies could be gradually removed (Moran et
al.\ 2007), but the additional mass such a reservoir could supply in
the short-term is probably insignificant compared to the large masses
already in the discs and cores. So, provided the molecular reservoirs
are securely bound to the galaxy, the cold gas mass we measure now is
likely to contribute to their final stellar masses. Their current
stellar masses (estimated from their $K$-band luminosities, see Geach
et al.\ 2009) are likely to be in the range $M_\star \sim
0.5$--$1.5\times10^{10}M_\odot$, implying that the current episode of
star formation could double their stellar mass (if the bursts are
half-way through their lifetime). Assuming most of the current star
formation occurs in the central regions, then relative fading of the
disc component compared to the maturing bulge will contribute to the
morphological transformation necessary for them to evolve into a
massive S0 (Kodama \& Smail\ 2001).

These galaxies appear to be undergoing significant evolution in the
outskirts of the cluster, with their activity potentially driven by
mergers and interactions. So how important is the central cluster
environment in transforming spirals to S0s, and maintaining the
evolution of the morphology-density relationship? Our observations
suggest that the cluster environment is not required to terminate star
formation in the progenitors of S0s; the key environmental effect
might be more subtle.  One can speculate on the likely processes that
the galaxies will experience as they are virialised (long after they
have used up their gas). First, the lack of accretion of additional
gas from a cooling halo will prevent additional star formation. It is
unlikely that the galaxies will experience further mergers, and so
they cannot acquire cold gas that way; all that remains is
morphological transformation. If the galaxies are on radial orbits,
then high speed passes in the cluster potential and close encounters
could promote tidal disruptions or even removal of any remaining
spiral disc (Moore, Lake\ \& Katz\ 1998; Gnedin\ 2003), completing the
transformation to S0.

\section{Summary}

We have presented new IRAM CO detections of two LIRGs in the outskirts
of the rich cluster Cl\,0024+16 at $z=0.395$. From the CO(1--0) line
luminosities, we measure H$_2$ gas masses of
$(5.4\pm0.5)\times10^{9}$\,$M_\odot$ and
$(9.1\pm0.9)\times10^{9}$\,$M_\odot$ in the two galaxies. This assumes
a conversion between CO luminosity and H$_2$ mass that is applied in
ULIRGs; if we adopt the Galactic conversion factor, these estimates
would increase by a factor $\sim$6$\times$. The infrared-derived SFRs
are $\sim$60\,$M_\odot$\,yr$^{-1}$, and so they will exhaust their
reservoirs in as little as $\sim$150\,Myr. 

Our observations hint that dusty starbursts in the outskirts of
Cl\,0024+16 can assemble $M_\star\sim 10^{10}M_\odot$ of stellar mass
long before they reach the cluster core, implying that the cluster
environment is unlikely to have a significant influence on the
burst. However, it is interesting to note that one of the galaxies
shows evidence of dynamic disturbance, potentially linked to a
merger. If the galaxies are destined to evolve into S0s, then the main
transformation required after the gas reservoirs have been exhausted
is fading of the disc relative to the bulge. Potentially this could be
accompanied by some morphological disturbance when the galaxies reach
the high density core.

On a final note, we highlight the fact that if the critical epoch of
star formation occurs $\sim$1--2\,Gyr before the galaxies reach the
virial radius, and these starbursts are feeding the local S0
population, then there should be a sizable population of
post-starburst (k+a) galaxies at $\sim$$R_{\rm vir}$ in rich clusters
at $z\sim0.2$--0.3 (Couch \& Sharples\ 1987). Detection of such an
`intermediate' population would be a key piece of evidence in the
model where starbursts triggered early during cluster infall at
high-redshift are rapidly exhausting their gas reservoirs and sinking
to the bottom of the cluster potential well where they are destined to
complete their evolution into local massive S0s.

\medskip

\noindent We warmly thank Roberto Neri and Philippe Salom\'e at IRAM
for their assistance in obtaining and reducing this data.  We also
thank Mark Swinbank for helpful comments. This work is based on
observations carried out with the IRAM PdBI. IRAM is supported by
INSU/CNRS (France), MPG (Germany) and IGN (Spain). JEG is supported by
the U.K. Science and Technology Facilities Council (STFC). KEKC is a
STFC Fellow at Durham University. IRS and RSE acknowledge the Royal
Society and STFC.

\label{lastpage}

\end{document}